\documentclass[aps,prr,reprint,superscriptaddress,longbibliography]{revtex4-2}

\usepackage{graphicx}
\usepackage{braket}
\usepackage{amsmath,amssymb,amsthm}
\usepackage{enumitem}
\usepackage{bbold}
\usepackage{xcolor}
\usepackage{natbib}
\usepackage[colorlinks=true,linkcolor=blue,citecolor=blue,urlcolor=blue]{hyperref}

\begin{document}

\title{Accelerated first detection in discrete-time quantum walks using sharp restarts}

\author{Kunal Shukla}
\email{kunalshukla@iisc.ac.in}
\affiliation{Dept. of Instrumentation \& Applied Physics, Indian Institute of Science, C.V. Raman Avenue, Bengaluru 560012, India}
\author{Riddhi Chatterjee}
\affiliation{Dept. of Instrumentation \& Applied Physics, Indian Institute of Science, C.V. Raman Avenue, Bengaluru 560012, India}
\author{C. M. Chandrashekar}
\affiliation{Dept. of Instrumentation \& Applied Physics, Indian Institute of Science, C.V. Raman Avenue, Bengaluru 560012, India}
\affiliation{The Institute of Mathematical Sciences, C. I. T. Campus, Taramani, Chennai 600113, India}
\affiliation{Homi Bhabha National Institute, Training School Complex, Anushakti Nagar, Mumbai 400094, India}


\begin{abstract}
Restart is a common strategy observed in nature that accelerates first-passage processes, and has been extensively studied using classical random walks. In the quantum regime, restart in continuous-time quantum walks (CTQWs) has been shown to expedite the quantum hitting times [\href{https://doi.org/10.1103/PhysRevLett.130.050802}{Phys. Rev. Lett. \textbf{130}, 050802 (2023)}]. Here, we study how restarting monitored discrete-time quantum walks (DTQWs) affects the quantum hitting times. We show that the restarted DTQWs outperform classical random walks in target searches, benefiting from quantum ballistic propagation, a feature shared with their continuous-time counterparts. Moreover, the explicit coin degree of freedom in DTQWs allows them to surpass even CTQWs in target detection without sacrificing any quantum advantage. Additionally, knowledge of the target's parity or position relative to the origin can be leveraged to tailor DTQWs for even faster searches. Our study paves the way for more efficient use of DTQWs in quantum-walk-based search algorithms, simulations and modeling of quantum transport towards targeted sites in complex quantum networks. 
\end{abstract}

\maketitle

\section{Introduction}
First-passage-time problems appear in diverse fields \cite{rednerGuideFirstPassageProcesses2001}, from ecology \cite{mendezRandomSearchStrategies2014} and genetics \cite{zhangFirstPassageProcessesGenome2016} to chemical kinetics \cite{aquinoChemicalContinuousTime2017}, finance \cite{chicheporticheApplicationsFirstPassageIdeas2013}, and disaster management \cite{wijesunderaFirstArrivalTime2016}, and involve questions such as when a predator finds prey, a genetic mutation first appears in a population, or a stock price hits a certain threshold. These problems can be modeled by classical random walks, which involve a random walker searching for a target \cite{metzlerFirstPassagePhenomenaTheir2014}. A common challenge is that the walker may stray from the target and revisit the same sites, indefinitely increasing the mean search time. Nature has a strategy to bypass this problem: restart the walk if the walker cannot find the target for some time \cite{rotbartMichaelisMentenReactionScheme2015,belSimplicityCompletionTime2009, reuveniRoleSubstrateUnbinding2014,munskySpecificityCompletionTime2009}. Employing this restart strategy has enhanced our classical search algorithms \cite{lubyOptimalSpeedupVegas1993} and has been extensively studied in relation to stochastic processes under restart \cite{evansDiffusionStochasticResetting2011, evansDiffusionOptimalResetting2011}. 

Classical random walks have a quantum counterpart known as the quantum walk, which comes in two forms: Continuous-time quantum walks (CTQW) \cite{venegas-andracaQuantumWalksComprehensive2012} and discrete-time quantum walks (DTQW) \cite{chandrashekarDiscreteTimeQuantumWalk2010}. The Hilbert space of the continuous-time variant consists of a position space spanned by sites on a lattice, while the discrete-time version includes a two-dimensional coin space alongside the position space. Both CTQWs and DTQWs have been applied to various problems \cite{farhiQuantumComputationDecision1998a, mulkenContinuoustimeQuantumWalks2011, lahiniAndersonLocalizationNonlinearity2008, razzoliLatticeQuantumMagnetometry2019, shuklaQuantumMagnetometryUsing2024, chawlaMultiqubitQuantumComputing2023}, from modeling energy transport in photosynthesis \cite{mohseniEnvironmentassistedQuantumWalks2008, engelEvidenceWavelikeEnergy2007} to simulating the Dirac equation \cite{mallickSimulatingDiracHamiltonian2019, chandrashekarTwocomponentDiraclikeHamiltonian2013}. Both types have also been experimentally implemented in multiple physical systems \cite{tornowMeasurementinducedQuantumWalks2023, dadrasExperimentalRealizationMomentumspace2019, ryanExperimentalImplementationDiscretetime2005, zahringerRealizationQuantumWalk2010, broomeDiscreteSinglePhotonQuantum2010, schreiberPhotonsWalkingLine2010}.

Unlike classical walks, we cannot have a complete record of the walker's trajectory in quantum walks as measurements collapse the quantum state, disrupting the quantum walker's evolution. Hence, instead of discussing first-passage time, we examine the first-detected-passage time (FDPT) \cite{friedmanQuantumWalksFirst2017, kroviQuantumWalksInfinite2006, thielFirstDetectedArrival2018, dharQuantumTimeArrival2015} at a target site. In classical walks \cite{palFirstPassageRestart2017}, first-detection and first-passage times at a target site are equivalent because one can frequently measure the walker's position without affecting the walk. In quantum walks, however, the measurement period significantly affects the walk dynamics. Previous studies \cite{yinInstabilityQuantumRestart2024, yinRestartExpeditesQuantum2023} on CTQWs evolving under a tight-binding Hamiltonian show that both the restart time and the measurement period are essential control parameters, contrary to classical walks where the latter is relatively inconsequential. While classical walks have a unique minimum mean first-passage time \cite{guptaStochasticResettingVery2022, evansDiffusionOptimalResetting2011, evansDiffusionResettingArbitrary2014, bonomoFirstPassageRestart2021}, CTQWs exhibit multiple minimum mean FDPTs under stroboscopic measurements. 

For DTQWs, recurrence without restart has been studied \cite{chenUnmonitoredMonitoredRecurrence2024, stefanakRecurrencePolyaNumber2008, stefanakRecurrencePropertiesUnbiased2008, xuDiscretetimeQuantumWalks2010}, but the combined effects of restart and measurement period on FDPTs remain unexplored. DTQWs provide better control over the walk dynamics than CTQWs due to their explicit coin degree of freedom. Studying DTQWs with restart is further motivated by their successful applications in quantum search algorithms \cite{shenviQuantumRandomwalkSearch2003,sahuQuantumwalkSearchMotion2024}, and implementations in physical systems \cite{senguptaExperimentalRealizationUniversal2024,ryanExperimentalImplementationDiscretetime2005,schreiberPhotonsWalkingLine2010,dadrasExperimentalRealizationMomentumspace2019,broomeDiscreteSinglePhotonQuantum2010}. In this work, we study the effect of restart on the FDPT of a quantum walker undergoing DTQW. We explore three main questions: How do DTQWs compare to CTQWs in target searching? Can the explicit coin degree of freedom in DTQWs reduce the mean FDPTs? Finally, what are the optimum restart times and measurement periods, and how do these depend on the target location? We will find that DTQWs outperform CTQWs, and their explicit coin degree of freedom can be exploited to expedite the FDPTs further. Moreover, knowledge of the target location's parity and position relative to the origin plays a crucial role in selecting the initial coin state of the walker, the restart time, and the measurement period. 

The paper is organized as follows: Section II introduces DTQWs and CTQWs models, with a primary focus on the sharp restart strategy and the measurement protocol used in this work. Section III provides the main results, addressing the three core questions posed. Finally, section IV closes the paper with the key results and possible outlooks. 

\section{Monitored quantum walks with sharp restart}

Let us begin by introducing DTQWs. Consider a one-dimensional lattice spanned by $2N + 1$ sites, labeled by integers from $-N$ to $N$. A quantum walker, such as an electron or a photon, with an internal degree of freedom (spin or polarization) undergoes DTQW over this lattice. The initial quantum state of the walker is represented by $\ket{\Psi(t=0)} = \ket{s} \otimes \ket{x = 0}$ existing in a Hilbert Space, $\mathcal{H} = \mathcal{C}\otimes\mathcal{H}_p$. Here, $\mathcal{C} = \text{span}\{\ket{0}, \ket{1}\}$ denotes the coin space, and $\mathcal{H}_p$ is the position space spanned by the lattice sites. The walk operator [$\hat{W} = \hat{S}(\hat{C} \otimes \mathbb{1}_{p})$] acts on the walker at fixed time intervals ($\varepsilon$), evolving its state in discrete time-steps. Consequently, after the $\eta^{th}$ step, the walker's quantum state is given by $\ket{\Psi(\eta\varepsilon)} = \ket{\psi_\eta} = \hat{W} \ket{\psi_{(\eta-1)}}$. The walk operator ($\hat{W}$) consists of the coin operator ($\hat{C}$) and the conditional shift operator ($\hat{S}$). In this work, we use the following coin operator:
\begin{equation}
    \hat{\textsf{C}}(\theta)= \exp{(-i \theta \sigma_{x})} = \begin{pmatrix}
        \cos(\theta) & -i\sin(\theta)\\
        -i\sin(\theta) & \cos(\theta)
        \label{eq: coin}
    \end{pmatrix},
\end{equation}
which rotates the spinor ($\ket{s}$) by an angle $2\theta$ about the positive x-axis. Throughout this work, coin parameter is fixed at $\theta = \pi/4$. The shift operator ($\hat{S}$) is defined as:
\begin{align}
    \hat{S} = \sum_{x} \Big [ \ket{0}\bra{0} \otimes \ket{x-1}\bra{x} + \ket{1}\bra{1} \otimes \ket{x+1}\bra{x}\Big ],
    \label{eq: shift} 
\end{align}
shifting the position state $\ket{x}$ associated with coin state $\ket{1}$ ($\ket{0}$) by one step in the positive (negative) direction. Note that this shift operator implements an unbounded DTQW on an infinite lattice with open boundary conditions.

Similarly, the  model we consider to implement CTQW and study sharp restart is the tight-binding evolution with repeated monitoring \cite{yinRestartExpeditesQuantum2023}. The quantum walker undergoing continuous-time evolution  moves on the same one-dimensional lattice as the discrete-time walker but evolves continuously under the tight-binding Hamiltonian given by
\begin{equation}\label{eq: tight_bindind_hamiltonian}
    H = -\gamma \sum_{x=-N}^{N}(\ket{x}\bra{x+1}+\ket{x+1}\bra{x}).
\end{equation}
The initial quantum state of the walker here is given by $\ket{\Phi(t=0)} = \ket{x = 0}$, which exists only in the position Hilbert space $\mathcal{H}_p$. The walker's state after time $\Delta t$ is given by $\Phi(t + \Delta t) = \exp(-iH\Delta t/\hbar)\Phi(t)$ with $\hbar=1$. Unlike a DTQW, where the walker evolves in discrete steps of time $\varepsilon$ (i.e., $\Psi(n\varepsilon)$), in a CTQW the time increment $\Delta t$ in the state $\Phi(t + \Delta t)$ can be chosen arbitrarily small, reflecting the continuous nature of the evolution.

To monitor the walks, we repeatedly measure the walker's position at $x = \delta$  after time intervals of $\Delta T$. In a CTQW, $\Delta T$ can be set arbitrarily small; however, in a DTQW, $\Delta T$ must be an integer multiple of the discrete time step $\varepsilon$. In other words, $\Delta T = \tau\varepsilon$ for DTQWs, where the integer $\tau$ represents the DTQW step just after which measurement is taken. For instance, $\tau = 2$ implies that we measure the walker's position after every two steps of DTQW. In both walks, each measurement gives a binary outcome --- \textit{yes}, the walker was found at $\ket{\delta}$ or \textit{no}, it was not. The walks continue as long as measurements return \textit{no} and stop upon the first \textit{yes} \cite{yinRestartExpeditesQuantum2023, kroviQuantumWalksInfinite2006, dharDetectionQuantumParticle2015, friedmanQuantumWalksFirst2017}. Consider the first measurement at time $\Delta T$. At time $\Delta T^- = \Delta T - \epsilon$ with $\epsilon \rightarrow 0$ being positive, the continuous-time walker's state is given by
\begin{equation}
    \ket{\Phi(\Delta T^-)} = \exp(-iH\Delta T)\ket{\Phi(0)}.
\end{equation}
Similarly, after choosing a suitable time step $\varepsilon$ such that $\Delta T$ ends up being the integer multiple of $\varepsilon$; the discrete-time quantum walker's state just before the first measurement is given by
\begin{equation}
    \ket{\Psi(\Delta T^- = \tau \varepsilon - \epsilon)} = \hat{W}^{\tau} \ket{\Psi(0)}. 
\end{equation}
After the first measurement, we detect the continuous-time quantum walker at $x = \delta$ with probability $p_1 = |\langle \delta|\Phi(\Delta T^-) \rangle|^2$, and the discrete-time quantum walker with probability $p_1 = \sum_{c = 0}^{1}|(\bra{c} \otimes \bra{\delta})\ket{\Psi(\Delta T^-)}|^2$. If the walker is detected, the walks end. However, if the walker is not detected, which occurs with probability $1 - p_1$, the quantum walks continues with a quantum state altered by the null measurement. The altered state just after a null measurement, i.e., at time $\Delta T^+ = \Delta T+\epsilon$ in the CTQW is given by
\begin{equation}
    \ket{\Phi(\Delta T^+)} = N(\mathbb{1}_p - \ket{\delta}\bra{\delta})\ket{\Phi(\Delta T^-)}
\end{equation}
where $\mathbb{1}_p$ is the identity operator on the position space and N is the normalisation constant. For DTQW, we simply replace $\mathbb{1}_p$ with $\mathbb{1}_c \otimes \mathbb{1}_p$ and $\ket{\delta}\bra{\delta}$ with $\mathbb{1}_c \otimes \ket{\delta}\bra{\delta}$, and of course $\Phi$ by $\Psi$.

We can restart the walks after every $r$ measurements, which corresponds to a time interval of duration $r \Delta T$. Immediately after every restart, the walker's quantum state is changed to the initial state, $\ket{\Psi(t = 0)}$ for DTQW, and $\ket{\Phi(t=0)}$ for CTQW, respectively.
In these monitored quantum walks under restart, we encounter three types of probabilities. The first is $p_n$ representing the probability of detecting the walker at $x=\delta$ on the $n$th measurement which we already discussed for $n=1$. The second, $F_n$, describes the probability of detecting the walker at $x = \delta$ for the first time on the $n$th measurement; $F_n = (1 - p_1)\dots(1-p_{n-1})p_n$. Lastly, the cumulative detection probability up to $n$ measurements is denoted by $P_{det}(n) = \sum_{k=1}^n F_k$. We are now ready to discuss the results.

\section{Results}
We begin by plotting $P_{det}(n)$ up to $n$ measurements (i.e., up to time $n\Delta T$) for four types of walks: in the quantum regime, CTQW and DTQW; in the classical regime, continuous-time classical random walk (CTRW) and discrete-time classical random walk (DTRW). The classical walks \cite{rednerGuideFirstPassageProcesses2001} are performed on the same discrete lattice as the quantum walks. The continuous-time random walk (CTRW) evolves continuously in time, while the discrete-time random walk (DTRW) evolves in discrete time steps (denoted by \(\varepsilon\)). Both walks are symmetric, meaning that the probability of hopping left or right is one-half. Following \cite{yinRestartExpeditesQuantum2023}, we set the measurement time period \(\Delta T = 0.25\) for both the quantum and classical continuous-time variants. For the discrete-time versions, we set \(\varepsilon = 0.25\) and \(\tau = 1\), so that \(\Delta T = \tau\varepsilon = 0.25\), ensuring that the time interval between consecutive measurements is \(0.25\) in all four walks. We use the symmetric initial spin state, i.e., \(\ket{s} = (\ket{0}+\ket{1})/\sqrt{2}\) for the DTQW, and set \(\gamma = 1\) [Eq.~\ref{eq: tight_bindind_hamiltonian}] for the CTQW. The target site \(\delta\) is set to 10 for all walks.

\begin{figure}[h]
    \centering
    \begin{minipage}{0.5\columnwidth}
        \centering
        \includegraphics[width=\linewidth]{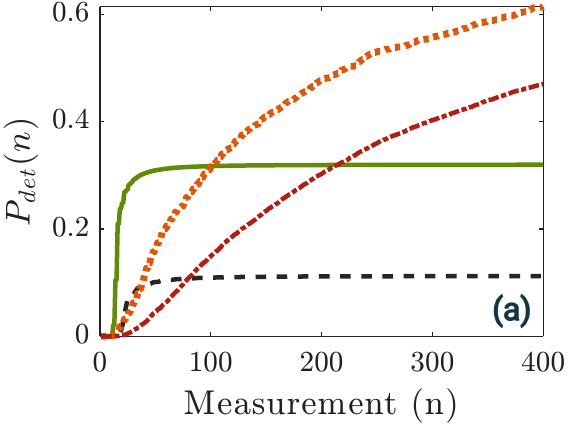}
    \end{minipage}\hfill
    \begin{minipage}{0.5\columnwidth}
        \centering
        \includegraphics[width=\linewidth]{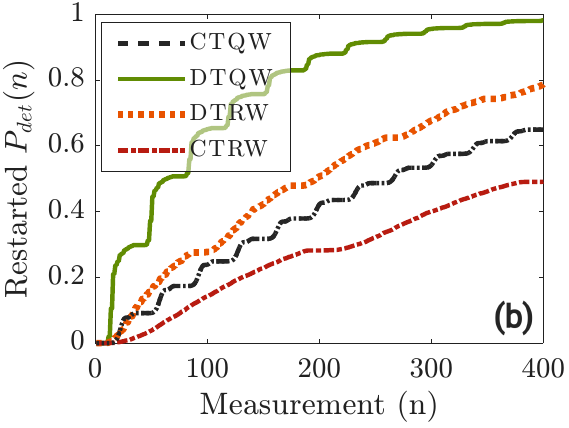}
    \end{minipage}
    \caption{(a) Cumulative detection probability $P_{det}(n)$ for discrete and continuous-time quantum and classical walks on an infinite lattice (open boundary conditions). While quantum walks initially outperform classical ones, at long times, their $P_{det}$ saturates at lower values, making quantum walks without restart worse than classical walks for target detection. (b) $P_{det}(n)$ for restarted walks, showing that discrete-time walks surpass the continuous-time counterparts, with DTQW achieving the fastest detection under restart. We set $\delta = 10$ for all the walks, $r = 35$ for quantum walks, and optimal restart times $r = 191$ and $87$ for CTRW, and symmetric DTRW, respectively. For discrete-time walks, $\varepsilon = 0.25$, $\tau = 1$; for continuous-time walks, $\tau = 0.25$.  Initial spin state $\ket{s} = \ket{+} = (\ket{0} + \ket{1})/\sqrt{2}$ is used for DTQW.}
    \label{fig: Pdet vs n for all walks}
\end{figure}

Figure \ref{fig: Pdet vs n for all walks}(a) shows $P_{det}(n)$ for all walks without restart. We see that DTQW outperforms all other walks in short times. However, at large times, classical walks beat quantum walks, as for classical walkers, $P_{det}(n)$ eventually reaches 1, and detection is guaranteed. In contrast, for quantum walkers, $P_{det}(n)$ remains significantly lower. Moreover, $P_{det}(n)$ of DTQW without restart saturates at a higher value than that of CTQW without restart.

Next, we employ the sharp-restart strategy \cite{palFirstPassageRestart2017,lubyOptimalSpeedupVegas1993} in which we restart the walks after every $r$ measurements. Keeping other parameters the same as before, we set $r = 35$ for quantum walks and the optimal restart times $r = 191$ \cite{yinRestartExpeditesQuantum2023} and $87$ for CTRW and DTRW, respectively, to ensure a fair comparison (Fig.~\ref{fig: Pdet vs n for all walks}(b)). In other words, we want to compare the quantum walks restarted at $r = 35$ (following \cite{yinRestartExpeditesQuantum2023}) with optimally restarted classical walks. We obtained the optimal restart time for DTRW for target site $\delta = 10$ by numerical simulation. We note two significant results. First, detection is now guaranteed for quantum walks as well: $P_{det}(n)$ of DTQW reaches 1 after around 400 measurements, significantly faster than other walks. In contrast, CTQW, with the set parameters (Fig.~\ref{fig: Pdet vs n for all walks}(b)), requires over 1000 measurements for its $P_{det}(n)$ to reach unity \cite{yinRestartExpeditesQuantum2023}. Second, while restarted CTQW outperforms optimally restarted CTRW \cite{yinRestartExpeditesQuantum2023}, it loses to restarted DTQW, and even to optimally restarted DTRW. Furthermore, restarted DTQW surpasses all other walks, highlighting its superiority in target detection.

\begin{figure*}[htbp]
    \centering
    \includegraphics[width=0.32\textwidth]{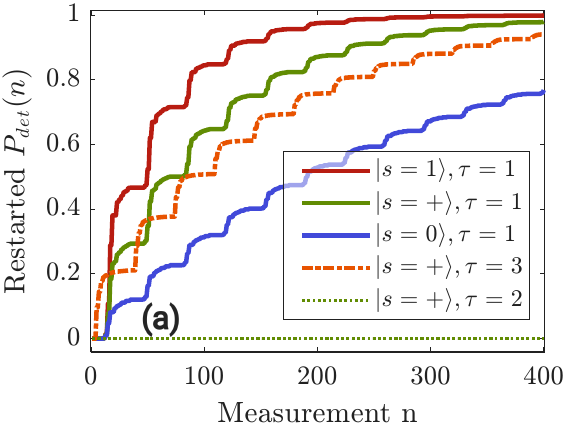}
    \includegraphics[width=0.32\textwidth]{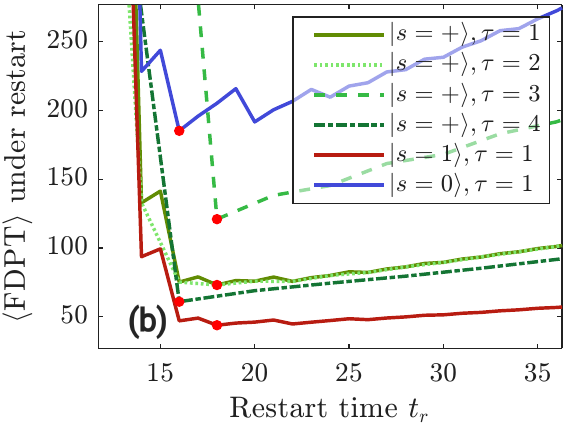}
    \includegraphics[width=0.32\textwidth]{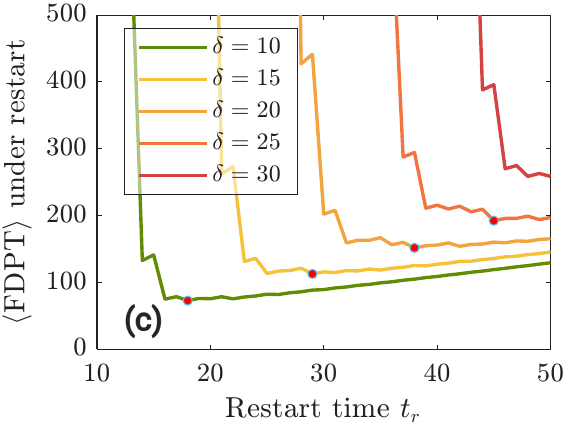}
    \caption{(a) $P_{det}(n)$ for restarted DTQWs, showing that when $\tau$ is constant and $\delta = 11 > 0$, initial spin state $\ket{1}$ performs the best, followed by $\ket{+}$ and $\ket{0}$, respectively. Plot also illustrates when $\delta$ is odd, even $\tau$s, such as $\tau = 2$, yield $P_{det}(n) = 0$ at every measurement. (b) Mean FDPT as a function of restart time $t_r$ for restarted DTQWs with constant $\delta = 10$ and varying $\tau$. (c) Mean FDPT versus $t_r$ with varying $\delta$ and constant $\tau = 1$; $\ket{s} = \ket{+}$.}
    \label{fig: Mean FDPTs}
\end{figure*}

We address two important questions before discussing further results: (1) Why does $P_{det}(n)$ for non-restarted quantum walks saturate at values below one? (2) Why do discrete-time walks, whether restarted or not, outperform continuous-time ones? The first question can be answered by considering the role of destructive interference, which is unique to quantum walks. In particular, under monitored measurements \cite{didiMeasurementinducedQuantumWalks2022}, quantum walks exhibit an undetectable dark subspace \cite{thielDarkStatesQuantum2020,pandharpatteDecidingFactorDetecting2024,thielUncertaintySymmetryBounds2020} that prevents $P_{det}(n)$ from reaching 1, even as $n$ increases indefinitely. In contrast, classical walks lack destructive interference because they operate with additive probabilities rather than probability amplitudes. Consequently, in classical walks, $P_{det}(n)$ eventually converges to 1 as $n$ increases. To address question (2), note that discrete-time walks incorporate an additional coin degree of freedom, which indirectly influences their position probability distributions (PPDs). However, the primary reason that discrete-time walks outperform continuous-time ones is the distinct nature of these PPDs:

\begin{enumerate}[label=\alph*.]
    \item \textbf{How PPDs help discrete-time walks:} In discrete-time walks, the detection probabilities, $p(x|\eta)$, are nonzero only at alternate lattice sites. Specifically, at even (odd) steps of the walk, the detection probabilities at odd (even) sites vanish, i.e., 
    \begin{equation}\label{eq: alternate}
        \begin{split}
            &p(x=\text{odd}|\eta=\text{even})=0\\
            &p(x=\text{even}|\eta=\text{odd})=0.
        \end{split}
    \end{equation}
    In contrast, continuous-time walks assign finite probabilities to all sites that the walker can reach at a given step. Since the total detection probability is conserved (i.e., sums to one), the probability of detecting the walker at $x=\delta$ at the $n$th measurement—denoted by $p_n$—is lower in continuous-time walks because the probability is distributed among more sites.
    
    \item \textbf{How PPDs help continuous-time walks:} For continuous-time walks, having finite $p(x|\eta)$ for every accessible site offers an advantage. At each measurement, the CTQW exhibits a nonzero detection probability $p_n$, whereas in discrete-time walks with an odd $\tau$ (e.g., $\tau=1$ in Fig.~1), $p_n$ vanishes at alternate measurements.
\end{enumerate}
Now, we compare the walks by evaluating their $P_{det}(n)$. The leading contribution in the expansion of $P_{det}(n)$ is $\sum_{i=1}^n p_i$, so the walk with the larger sum performs better. In other words, points (a) and (b) compete. The $P_{det}(n)$ plots in Fig.~\ref{fig: Pdet vs n for all walks} show that point (a) prevails; that is, the advantage of higher $p_n$ values in discrete-time walks outweighs the benefit of having finite $p_n$ at every $n>\delta$ in continuous-time walks, leading to better overall performance for the discrete-time walks. 

We now focus on restarted DTQWs to study how parameters such as initial spin state ($\ket{s}$) and measurement step ($\tau = \Delta T/\varepsilon$) affect quantum hitting times. In Fig.~\ref{fig: Mean FDPTs}(a), we plot $P_{det}(n)$ over $n$ measurements for DTQWs with different $\tau$ values and initial spin states $\ket{s}$. Although the plots share the same measurement index (the x-axis), the actual measurement times ($n\Delta T$) differ. For example, for walks with $\tau=1$ and $\tau=3$, $P_{det}(n)$ corresponds to cumulative detection probabilities up to times $n\varepsilon$ and $3n\varepsilon$, respectively. We see that for the target positions $\delta > 0$ (here, $\delta =11$), initial spin state $\ket{s} = \ket{1}$ that produces the left-skewed PPDs performs the best. In contrast, $\ket{s} = \ket{0}$ that produces right-skewed PPDs performs the worst. Thus, we conclude that when the target's position relative to the origin is known, the search can be expedited by choosing an initial state that skews the PPD toward the side of the origin opposite the target. In other words, for the chosen coin [Eq.~\ref{eq: coin}] and shift operator [Eq.~\ref{eq: shift}], choose $\ket{s} = \ket{0}(\ket{s}=\ket{1})$ when $\delta<0(\delta>0)$. When such information is unavailable, setting $\ket{s} = \ket{+}$ that produces symmetric PPD is the safer choice. We will talk about why this happens in some detail later. Fig.~\ref{fig: Mean FDPTs}(a) also shows that when $\delta$ is odd (here, $\delta = 11$), the walker can never detect the target if $\tau$ is chosen to be even (here, $\tau = 2$). While odd $\tau$s allow detection for both even and odd $\delta$s, detection is impossible for odd $\delta$ when even $\tau$s are used. The reason for this behavior is examined below.

In DTQWs, the probability of finding the walker at all even (odd) positions is 0 at odd (even) steps of the walk [Eq.~\ref{eq: alternate}]. While for odd $\tau$s (say $\tau = 3$), we measure the walker's position at both odd and even steps alternatively (at $\eta =3, 6, 9, ...$), for even $\tau$s, we can measure position only at even steps. Thus, for even $\tau$s, detection at odd $\delta$ is impossible. These observations not only suggest that setting an odd $\tau$ is a safer option when target's parity is unknown, but also that if it is known, setting an even $\tau$ could be more beneficial. This is because when $\delta$'s parity is known, it can always be turned even by shifting the origin. Then, for even $\tau$s, we have finite detection probability at every measurement step, whereas for odd $\tau$s, we definitely miss detection at every alternate odd measurement step. This will be demonstrated later, with some nuances. However, first, let us move from $P_{det}(n)$ to expected FDPT under restart, denoted by $\langle\text{FDPT}\rangle_r$, which is a more robust measure of the efficiency of target detection.

Let $n_{f}$ be the number of measurements, and $\mathcal{R}$ be the number of restarts until first hitting. If we restart, as before, after every $r$th measurement step, $\langle n_f \rangle = r \langle \mathcal{R}\rangle + \langle\tilde{n}\rangle$, where $0 \le \mathcal{R} < \infty$ and $1 \le \tilde{n} \le r$. From here onward, we set the discrete time step $\varepsilon=1$ (without loss of generality), thereby simplifying the relation to $\langle\text{FDPT}\rangle_r = \Delta T \langle n_f \rangle =  \tau \langle n_f \rangle$ where \cite{eliazarTailbehaviorRoadmapSharp2021, bonomoFirstPassageRestart2021, lubyOptimalSpeedupVegas1993}

\begin{equation} \label{eq: ind_noise_model}
    \begin{split}
        \langle n_f \rangle &= \sum_{\mathcal{R} = 0}^{\infty} \sum_{\tilde{n} = 1}^{r} (r \mathcal{R} + \tilde{n})[(1-P_{det}(r))^\mathcal{R}F_{\tilde{n}}] \\
        & = \frac{1 - P_{det}(r)}{P_{det}(r)}r + \sum_{\tilde{n} = 1}^{r}\frac{(\tilde{n})F_{\tilde{n}}}{P_{det}(r)}.
    \end{split}
\end{equation}
The final equality follows from derivations for CTQW \cite{yinRestartExpeditesQuantum2023}.

In Fig.~\ref{fig: Mean FDPTs}(b), we plot $\langle\text{FDPT}\rangle_r$ as a function of restart time $t_r = r\Delta T = r\tau (\because \varepsilon=1)$, keeping $\delta = 10$ fixed for all walks. We see an oscillatory behavior in the mean FDPTs explicit to quantum walks. The results also support the claim that DTQWs initialized with a spin state $\ket{s}$ that skews the PPD toward the side opposite the target outperform those with other initial states. Specifically, for $\tau = 1$, the average first detection passage time, $\langle\text{FDPT}\rangle_r$, is lowest when $\ket{s} = \ket{1}$ and highest when $\ket{s} = \ket{0}$. This occurs because the target site $\delta = 10$ is on the right (positive) side of the origin, and initializing with $\ket{s} = \ket{1}$ produces a left-skewed PPD—i.e., the peak of the PPD is closer to the target. Consequently, the probability of detecting the walker at $\delta = 10$ is higher for DTQWs initialized with $\ket{s} = \ket{1}$ compared to those with $\ket{s} = \ket{0}$, where the PPD peaks away from the target. The increased detection probability for $\ket{s} = \ket{1}$ reduces the FDPT on average, as reflected in Fig.~\ref{fig: Mean FDPTs}(b). When $\ket{s}$ is fixed at $\ket{+}$, $\tau = 4$ yields the lowest expected FDPT under restart, followed closely by $\tau = 1$ and $2$, with $\tau = 3$ performing worst among them. We will compare the efficiency of different $\tau$s in detail later. 

Now, we note another interesting result: for the same $\delta$ ($10$) and $\ket{s}$ ($\ket{+}$), $\tau = 2$ closely bounds the $\langle\text{FDPT}\rangle_r$ plot of $\tau = 1$ from below. More generally, for all even $\delta$s, $\tau = 2\tau'$ \emph{tightly bounds} $\tau'$ whenever $\tau'$ is odd. This is expected when we compare the evolution of the DTQW for an odd $\tau'$ with its immediate even successor $2\tau'$. We find that for odd $\tau'$, where we measure walker's position at $x = \delta$ at times $n\Delta T = \tau',2\tau',3\tau',$ and so on. Since, $\delta$ is set even (here, $\delta = 10$), measurements at the alternate odd steps ($\tau', 3\tau', 5\tau',...)$ do not alter the walker's quantum state, making both walks essentially the same. The minor difference arises due to their PPD following the equation: $p_{2\tau'}(x = \delta | n) = p_{\tau'}(x = \delta | 2n)$. Here, $p_{2\tau'}$ and $p_{\tau'}$ is the probability of detecting the walker at $\delta$ during the $n$th and $2n$th measurement (i.e., effectively at same times $t_1 = n \Delta T_1 = n(2\tau')= 2n(\tau') = 2n\Delta T_2 = t_2$)  for DTQWs with measurement periods $2\tau'$ and $\tau'$, respectively. We conclude that, if the target location’s parity is known, selecting an even $2\tau'$ is preferable, as using the odd half $\tau'$ results in redundant measurements at every alternate step.

Before finding the most efficient $\tau$, let us examine how the expected FDPT under restart varies with $\delta$. In Fig.~\ref{fig: Mean FDPTs}(c), keeping constant $\tau=1$, we plot $\langle\text{FDPT}\rangle_r$ as a function of restart time $t_r= r\Delta T = r\tau$ for different $\delta$s. In both Fig.~\ref{fig: Mean FDPTs} (b) and (c), the y-coordinate of the red points indicate minimum $\langle\text{FDPT}\rangle_r$, while their x-coordinate represent the optimum restart time. The optimum restart time (ORT), i.e., the restart time that minimizes $\langle\text{FDPT}\rangle_r$, increases as $\delta$ grows. One expects this result because to reach $\delta$ far from the origin, the walker needs to evolve for a longer time, necessitating a larger restart time. However, this increase in ORT with $\delta$ may not be strictly monotonic.

\begin{figure*}[htbp]
    \centering
    \includegraphics[width=0.32\textwidth]{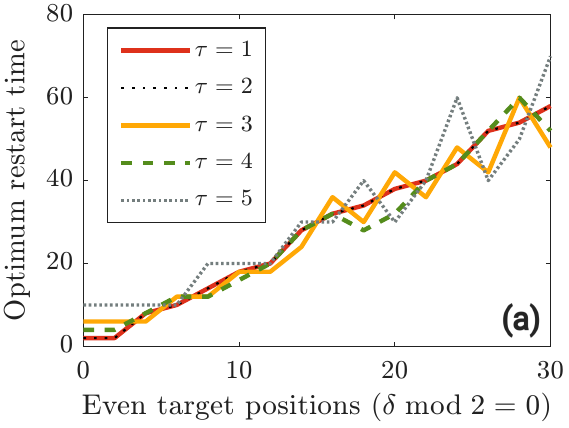}
    \includegraphics[width=0.32\textwidth]{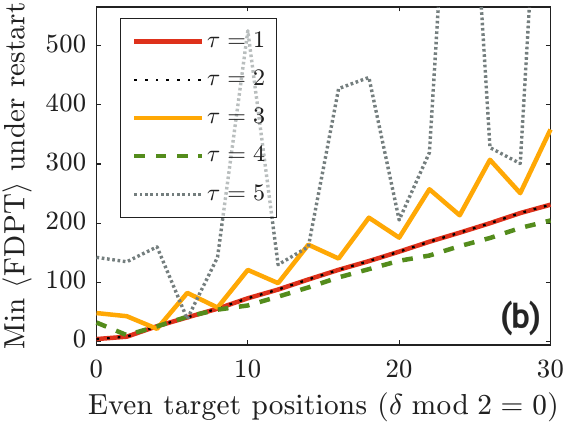}
    \includegraphics[width=0.32\textwidth]{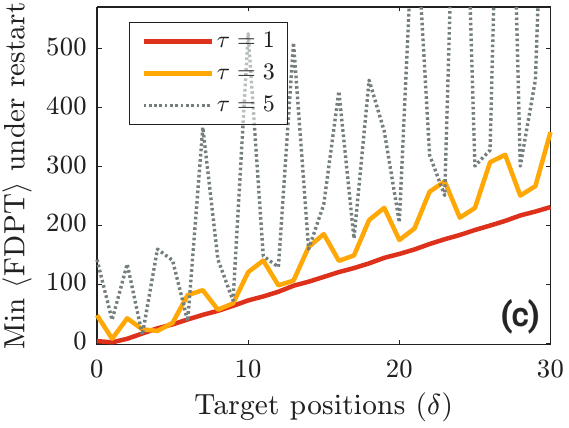}
    \caption{(a) Optimum restart times as a function of even values of $\delta$. Plots, except $\tau = 1$ and $2$, exhibit a non-monotonic increase as $\delta$ grows; however, they all maintain a positive average slope. (b) Minimum values of $\langle \text{FDPT} \rangle$ versus even $\delta$s, showing that for even target positions, $\tau = 4$ is the optimal measurement period. (c) When min $\langle \text{FDPT} \rangle$ is plotted for all $\delta$ values (even \& odd), only odd $\tau$s can compete, with $\tau = 1$ providing the fastest search. All plots correspond to restarted DTQWs initialized with $\ket{s} = \ket{+}$.}
    \label{fig: ORT and min mean FDPT}
\end{figure*}

To illustrate this, we plot ORT as a function of even $\delta$s in Fig.~\ref{fig: ORT and min mean FDPT}(a). We keep only even $\delta$s to allow even $\tau$s along with the odd ones. For odd $\delta$s and even $\tau$s, detection is impossible, causing ORT to diverge to infinity. The plot shows that for $\tau=1$, ORT increases monotonically with $\delta$, as also observed in Fig.~\ref{fig: Mean FDPTs}(c). The $\tau = 2$ plot lies precisely on the top of $\tau = 1$, further demonstrating the $\tau' - 2\tau'$ equivalence for even $\delta$s. As $\tau$ increases from $3$ to $5$, sharper peaks and valleys emerge in the plots. However, the average slopes of all the $\tau$ values are similar, indicating that as $\delta$ increases, ORT generally increases as well. Finally, we examine how varying $\tau$ affects the search for different target positions to identify the optimal measurement period.  

We divide the search for an optimal $\tau$ into two cases: (1) when the parity of the target position is known, and (2) when it is unknown. Case (1) can be simplified to a search for even $\delta$s in which both even and odd measurement periods can compete. In contrast, for case (2), we must only use odd $\tau$ values to ensure detection for all $\delta$s. Fig.~\ref{fig: ORT and min mean FDPT}(b) and (c) show the minimum expected FDPT as a function of $\delta$s for cases (1) and (2), respectively. The plots reveal that for even target positions, $\tau = 4$ yields the smallest values of minimum $\langle\text{FDPT}\rangle_r$ fiercely competing with $\tau = 1$ and $2$ whose plots, as expected, are identical. All values of $\tau > 2$ (except $\tau =4$) fall far behind in this race of target search. Hence, when the parity of the target position is known, we can turn it into an even value by shifting the origin if required. Then $\tau = 4$ is the optimal measurement period that gives the smallest values of minimum $\langle\text{FDPT}\rangle_r$, enabling the fastest search. In case (2), where we search for a target of unknown parity, Fig.~\ref{fig: ORT and min mean FDPT}(c) reveals that $\tau = 1$ is optimal. We now turn to the reasons behind these observations.

To explain these results, we examine the expression for $\langle\text{FDPT}\rangle_r$ as a function of $\Delta T$. Keeping $\varepsilon=1$ as before such that $\Delta T=\tau$, we have
\begin{equation}
    \begin{split}
        \langle\text{FDPT}\rangle_r(\tau) &= \tau\langle n_f \rangle \\
        &=  \frac{1 - P_{det}(r)}{P_{det}(r)}(r\tau) + \sum_{\tilde{n} = 1}^{r}\frac{(\tilde{n})F_{\tilde{n}}}{P_{det}(r)}\tau.
    \end{split}
    \label{eq: Fn}
\end{equation}
Here, $F_{\tilde{n}} = (1-p_1)...(1-p_{\tilde{n} -1})p_{\tilde{n}}$, as defined earlier. The first-order term with the highest contribution in the expansion of $F_n$ is $p_n$, followed by the second-order term, $p_n\sum_{i=1}^{n-1} p_n$, and so forth. However, approximating $F_n$ as $p_n$ is sufficient to explain the observed results. Similarly, the leading contribution in the expansion of $P_{det}(n)$ is $\sum_{i=1}^n p_i$, thus we approximate $P_{det}(n) \sim \sum_{i=1}^n p_i$.

Why do walks with high $\tau$ values perform poorly compared to those with lower $\tau$s? In other words, why walks with higher $\tau$ exhibit higher minimum $\langle\text{FDPT}\rangle_r$ values (see Fig.~\ref{fig: ORT and min mean FDPT}(b) and (c))? This occurs because a larger $\tau$ allows the quantum walker to spread further across the lattice between measurements. For example, after 2 null measurements, a DTQW walker with $\tau=5$ reaches $x=2\times5=10$, which is five times farther than one with $\tau=1$, reaching $x=2$. Hence, in high $\tau$ walks, compared to those with smaller $\tau$, the finite, conserved total probability is distributed over a larger domain, causing $p_n$ to take smaller values even with restarts. Now, keeping $\tau$ constant, if we minimize Eq.~\ref{eq: Fn} with respect to $r$, we obtain the minimum $\langle\text{FDPT}\rangle_r$ for a given $\tau$. The restart time $r\tau$ after minimization becomes the ORT, $r^*\tau$, which remains nearly the same for walks with different $\tau$ (see Fig.~\ref{fig: ORT and min mean FDPT}(a)). Moreover, setting $r=r^*$, and noting that $F_n\sim p_n$ and $P_{det}(n)\sim\sum_{i=1}^n p_i$, a straightforward analysis shows that both terms in Eq.~\ref{eq: Fn} increase with $\tau$, implying that the minimum $\langle\text{FDPT}\rangle_r$ rises as we increase $\tau$. In summary, walks with higher $\tau$ perform poorly compared to those with lower $\tau$.

However, if higher $\tau$s, which spread probability more widely across the lattice, generally increase $\langle\text{FDPT}\rangle_r$, then how does $\tau = 4$ outperforms $\tau = 1,2,$ and $3$ for even $\delta$ values? To answer this, two key points are relevant: (1) In general DTQWs, including those in this study, the probability of finding the walker at the position $x_{end}^{\pm}$ is very small. Here, $x_{end}^{\pm}$ represents the farthest lattice point with a non-zero chance of finding the walker; (2) Monitored-DTQW with $\tau = 1$ or $2$ and $\delta > 0$ ($\delta < 0$) has $x_{end}^+ \le \delta$ ($x_{end}^- \ge \delta$). Alternatively, assuming $\delta > 0$, the first time probability of measuring the walker at $x=\delta$ becomes non-zero, a null measurement changes the walker's quantum state such that $x_{end}$ turns from $\delta$ to $\delta - 2$. Thereafter, with only $1$ or $2$ steps between measurements ($\tau = 1$ or $2$), the walker can evolve only as far as $x_{end}^+ = \delta$, where it again faces the same outcome.

Points (1) and (2) explain why, for even $\delta$ values, $\tau = 4$ outperforms $\tau = 1$ and $2$, despite the broader probability spread. With $\tau = 4$, the walker moves beyond $\delta$, allowing its $p_n$ to exceed the limited $p_n = p(x_{end}^{\pm})$ achievable by $\tau = 1$ or $2$. Contrastingly, walks with $\tau = 3$ lose to those with $\tau = 1,2$, and $4$, even after allowing its walker to go past $\delta$. This is due to redundant alternate measurements that make no change in the walker's state, effectively allowing 6-step evolution and resulting in a broader probability spread that counteracts the benefits of moving past $\delta$.

We note that the results here are obtained for the coin $\hat{\textsf{C}}(\theta=\pi/4)$ [Eq.~\ref{eq: coin}]; however, they fundamentally depend on the nature of PPDs produced by the DTQWs. Consequently, with appropriate modifications, these results are generic and remain valid even if a different coin (e.g., the Hadamard coin) is used, provided that the analysis accounts for the corresponding PPDs. \\

\section{Conclusion}
In this work, we explored the role of sharp-restarts, initial states, and measurement periods in expediting the quantum hitting times of DTQWs. First, we highlight the key findings: Monitored DTQWs with sharp restart significantly outperform their continuous-time quantum counterpart and optimally restarted continuous-time and discrete-time classical walks. As shown in restarted CTQWs \cite{yinRestartExpeditesQuantum2023}, restart helps in evading dark states in the DTQWs context as well. The coin degree of freedom can be leveraged to tailor DTQWs with faster hitting times when the target's position relative to the origin is known (i.e., $\delta \ge 0$ or $\delta \le 0$). Notably, the mean FDPTs of restarted DTQWs exhibit oscillatory behavior, a hallmark of quantum behavior absent in classical walks. For higher measurement periods, the optimum time to restart the walk does not increase in a strictly monotonic fashion as the target moves farther from the origin. Furthermore, knowledge of the target's parity can be utilized to get an even more efficient search by selecting measurement period $\tau = 4$. The combined knowledge of the target's side and parity enables highly efficient searches. Even without such prior information, initializing the walk with a symmetric coin state and setting the measurement period to one still ensures better performance than classical and continuous-time quantum walks. 

With the ability to control and engineer periodic \cite{kumarBoundsDynamicsPeriodic2018a} and aperiodic DTQW \cite{logulloDynamicsEnergySpectra2017}, one can foresee an important role of sharp restart in modeling and accelerating quantum dynamics in complex quantum systems and quantum networks with directed targets.  


\bibliography{restartbetter1}

\end{document}